\documentclass[aps,showpacs,superscriptaddress,preprint]{revtex4}%
\usepackage{amsfonts}
\usepackage{amsmath}
\usepackage{amssymb}
\usepackage{graphicx}%
\begin{document}
\title{The effect of the warping term on the fractional quantum Hall states in
topological insulators}
\author{Fawei Zheng }
\affiliation{LCP, Institute of Applied Physics and Computational Mathematics, P. O. Box
8009, Beijing 100088, China}
\author{Zhigang Wang}
\affiliation{LCP, Institute of Applied Physics and Computational Mathematics, P. O. Box
8009, Beijing 100088, China}
\author{Zhen-Guo Fu}
\affiliation{Beijing Computational Science Research Center, Beijing
100084, China}
\author{Ping Zhang}
\thanks{Corresponding author. Email address: zhang\_ping@iapcm.ac.cn}
 \affiliation{LCP, Institute of Applied Physics and
Computational Mathematics, P. O. Box 8009, Beijing 100088, China}
\affiliation{Beijing Computational Science Research Center, Beijing
100084, China}


\begin{abstract}
The warping effect on the fractional quantum Hall (FQH) states in topological
insulators is studied theoretically. Based on the perturbed wavefunctions,
which include contributions from the warping term, analytical expressions for
Haldane's pseudopotentials are obtained. We show that the warping term does
not break the symmetry of the pseudopotentials for $n$=$\pm1$ Landau levels
(LLs). With increasing the warping strength of the Fermi surface, our results
indicate that the stability of the FQH states for LL $n=0$ (LLs $n$=$\pm1$)
becomes stronger (weaker), and the excitation gap at $\nu=1/3$ FQH state for
LL $n$=$0$ also increases while the gaps for LLs $n$=$\pm1$ are unchanged.

\end{abstract}
\maketitle

Topological insulators (TIs) as a new phase of quantum matter, which can not
be adiabatically connected to conventional insulators and semiconductors, have
been intensively studied in recent years \cite{Hasan, Qi, Bernevig, Fu,
Konig,Hsieh, ZhangHJ,Xia,Chen, Hsieh2009}. TIs are characterized by a full
insulating gap in the bulk and protected gapless edge or surface states. Near
the Fermi level the low-energy dispersion of the TI surface states shows a
Dirac linear behavior. However, the recent angle-resolved photoemission
spectroscopy experiments show that the Fermi surface in Bi$_{2}$Te$_{3}$
\cite{Chen,Hsieh2009}, a typical TI, is a snowflake shape rather than a circle
one. The origin of this snowflake-like Fermi surface has been confirmed to
arise from an unconventional hexagonal warping term \cite{Fu2009}. It is this
warping term that brings about many unique physical phenomena \cite{ZhangT,
WangJ,Xue}, which can not be observed in other systems, including the
extensively studied graphene and the conventional two-dimensional electron gas.

More recently, there has been emerging attention to the interactions of the
Dirac-type quasiparticles and their strong correlation effects in TI,
especially the TI fractional quantum Hall (TIFQH) states. Despites no
undeniable experimental observation of the TIFQH states heretofore, some
theoretical studies have been undertaken. For example, DaSilva \cite{DaSilva}
predicted the stability of the TIFQH states for Landau levels (LLs) with index
$n$=$0$ and $\pm1$ in TIs. Apalkov and Chakraborty studied the finite
thickness effect on the TIFQH states \cite{Chakraborty}. Also, the present
authors investigated the influences of the Zeeman splitting and the tilted
strong magnetic field on the stability of the TIFQH states \cite{Wang,Zheng}
with large $g$ factor. However, many important and interesting open questions,
such as the warping effect, the spin excitations, and the subband-LL coupling,
have not been discussed in the TIFQH regime.

In this paper we theoretically study the warping effect on the TIFQH states.
Here the warping term is perturbativly treated. With the help of the numerical
calculations, we show that the warping term can not break the symmetry of the
Haldane's pseudopotentials for $n$=$\pm1$ LLs, which differs from the role of
the spin splitting \cite{Wang}. Moreover, our results indicate that with the
increase of the warping strength of the Fermi surface, the stability of the
TIFQH states for LL $n=0$ (LLs $n$=$\pm1$) become stronger (weaker), and the
excitation gap at $\nu=1/3$ filling for LL $n$=$0$ (gaps for LLs $n$=$\pm1$)
also increases (keep unchanged).

In the presence of a perpendicular magnetic field $\mathbf{B}=B\hat{z}$, the
effective Dirac Hamiltonian for Bi$_{2}$Te$_{3}$(111) surface is written as
\begin{equation}
H_{0}=v_{f}\left(  \sigma_{x}\Pi_{y}-\sigma_{y}\Pi_{x}\right)  +\frac{\lambda
}{2}\left(  \Pi_{+}^{3}-\Pi_{-}^{3}\right)  \sigma_{z}, \label{1}%
\end{equation}
where $\mathbf{\Pi}=\mathbf{k}+e\mathbf{A}/c$ with the wave vector
$\mathbf{k}=\left(  k_{x},k_{y},0\right)  $ and the gauge $\mathbf{A}%
=B(-y/2,x/2,0)$. Here, $\Pi_{\pm}=\mathbf{\Pi}_{x}\pm\mathbf{\Pi}_{y}$,
$\sigma_{x,y,z}$ are Pauli matrices, $v_{f}$ denotes the Fermi velocity, and
$\lambda$ describes the hexagonal warping strength of the Fermi surface
\cite{Fu2009}. Here we have assumed that the Zeeman splitting is much weaker
than the warping term and therefere can be neglected for the first step in
order to solely illustrate the role played by the warping term. This is the
case for Bi$_{2}$Te$_{3}$(111) system. By introducing the ladder operators
$a^{\dag}=\frac{1}{\sqrt{2}}\left(  \frac{z}{2l_{B}}-2l_{B}\partial_{\bar{z}%
}\right)  $ and $a=\frac{1}{\sqrt{2}}\left(  \frac{\bar{z}}{2l_{B}}%
+2l_{B}\partial_{z}\right)  $ with $z(\bar{z})$=$x\pm iy$ and the magnetic
length $l_{B}=\sqrt{\hbar c/eB}$, we can rewrite the Hamiltonian (\ref{1}) as%
\begin{equation}
H_{0}=\frac{\sqrt{2}}{l_{B}}\left(
\begin{array}
[c]{cc}%
-i\left(  a^{\dag3}-a^{3}\right)  \lambda/l_{B}^{2} & v_{f}a\\
v_{f}a^{\dag} & i\left(  a^{\dag3}-a^{3}\right)  \lambda/l_{B}^{2}%
\end{array}
\right)  . \label{H0}%
\end{equation}
When $\lambda$=$0$ the Hamiltonian (\ref{H0}) can be exactly solved, and the
eigenstates are given by
\begin{equation}
\Psi_{n,m}^{(0)}=\left\{
\begin{array}
[c]{cc}%
\frac{1}{\sqrt{2}}\left(
\begin{array}
[c]{c}%
\text{sgn}(n)||n|-1,m\rangle\\
||n|,m\rangle
\end{array}
\right)  , & \text{ for }n\neq0,\\
\left(
\begin{array}
[c]{c}%
0\\
|0,m\rangle
\end{array}
\right)  , & \text{for }n=0,
\end{array}
\right.  \label{eigvectorH0}%
\end{equation}
where the symbol $|n,m\rangle$ is the non-relativistic two-dimensional
electron gas Landau eigenstates with non-relativistic quadratic dispersion
relation in $n$th LL with angular momentum $m$. The corresponding LLs are
expressed as $\varepsilon_{n}=sgn\left(  n\right)  v_{f}\sqrt{2\left\vert
n\right\vert }/l_{B}$. When the warping term is taken into account
($\lambda\neq0$), however, the single-particle eigenstate can not be obtained
directly. Fortunately, one can take the perturbation method to get eigenstates
of Hamiltonian (\ref{H0}) since the warping term, $\sqrt{2}\lambda l_{B}^{-3}%
$, is much smaller than the typical energy space between the
nearest-neighboring LLs, $\sqrt{2}v_{f}/l_{B}$, i.e., $\zeta\equiv
\lambda/(v_{f}l_{B}^{2})\ll1$. After some long but straightforward algebraic
operations, and only keeping the first-order terms, we have
\begin{equation}
\Psi_{n,m}=\Psi_{n,m}^{(0)}+\left(
\begin{array}
[c]{c}%
\chi_{|n|+2}^{(1)}||n|+2,m\rangle\\
\chi_{|n|-3}^{(2)}||n|-3,m\rangle
\end{array}
\right)  , \label{state1}%
\end{equation}
where the coefficients are%
\begin{align}
\chi_{|n|+2}^{(1)}  &  =-i\zeta\sqrt{\frac{\left(  |n|+1\right)  (|n|+2)}{2}%
},\nonumber\\
\chi_{|n|-3}^{(2)}  &  =\left\{
\begin{array}
[c]{c}%
-i\text{sgn}(n)\zeta\sqrt{\frac{\left(  |n|-1\right)  (|n|-2)}{2}}\\
0
\end{array}%
\begin{array}
[c]{c}%
|n|>2\\
|n|\leqslant2
\end{array}
\right.  .
\end{align}

\begin{figure}[ptb]
\begin{center}
\includegraphics[width=0.6\linewidth]{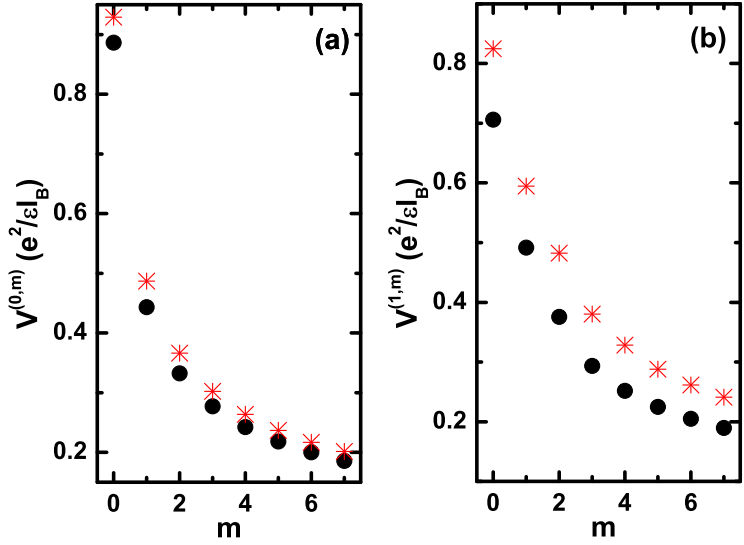}
\end{center}
\caption{(Color online) The effective pseudopotentials of the Coulomb
interaction $V^{(n,m)}$ between two electrons at (a) $n$=$0$ and (b) $n$=$1$
LLs as functions of the relative angular momentum with different warping
strength $\zeta$=$0$ (circles) and $\zeta$=$0.2$ (stars).}%
\label{Fig1}%
\end{figure}\begin{figure}[ptbptb]
\begin{center}
\includegraphics[width=0.6\linewidth]{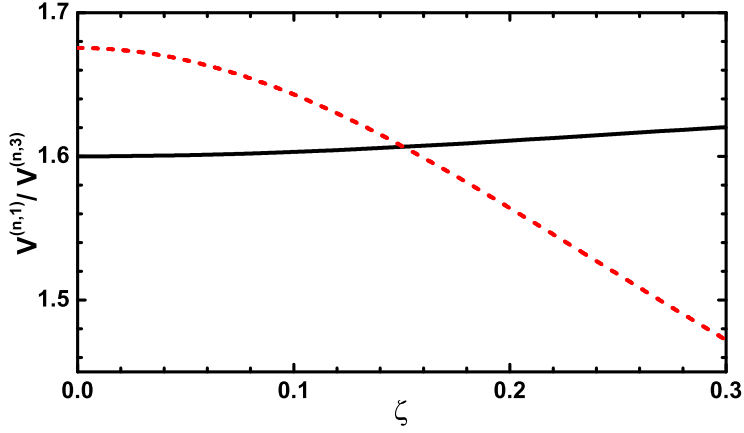}
\end{center}
\caption{(Color online) The ratio of the first and third relative angular
momentum pseudopotentials for LLs $n$=$0$ (black solid line) and $n$=$\pm1$
(red dashed line) as a function of the warping strength $\zeta$.}%
\label{Fig2}%
\end{figure}In the following discussion we will focus our attention to the
TIFQH states of $|n|\leqslant1$, because the stable TIFQH states can only be
observed for LLs $|n|\leqslant1$ \cite{DaSilva}. The Haldane's pseudopotential
\cite{Haldane} for Coulomb interaction $V(\mathbf{r})=\frac{e^{2}}{\epsilon
r}$ between electrons in the $n$th LL with relative angular momentum $m$ is
given by%
\begin{equation}
V^{\left(  n,m\right)  }=%
%TCIMACRO{\dsum \limits_{\mathbf{q}}}%
%BeginExpansion
{\displaystyle\sum\limits_{\mathbf{q}}}
%EndExpansion
\frac{\pi e^{2}}{\epsilon q}\left[  \mathcal{F}(q)\right]  ^{2}L_{m}%
(q^{2}l_{B}^{2})e^{-\frac{q^{2}l_{B}^{2}}{2}},
\end{equation}
in terms of Laguerre polynomials $L_{m}(x)$ and the form factor $\mathcal{F}%
(q)=\langle\Psi_{n}|e^{-i\mathbf{q}\cdot\mathbf{\eta}}|\Psi_{n}\rangle$ with
the cyclotron variable $\mathbf{\eta}=\mathbf{r}-\mathbf{R}$. Here,
$\mathbf{R}$ is the guiding-center position. Explicitly, for LLs
$|n|\mathtt{\leqslant}1$ we have
\begin{equation}
V^{\left(  0,m\right)  }=%
%TCIMACRO{\dsum \limits_{\mathbf{q}}}%
%BeginExpansion
{\displaystyle\sum\limits_{\mathbf{q}}}
%EndExpansion
\frac{\pi e^{2}}{\epsilon q}L_{m}(2x)e^{-2x}\left[  L_{0}\left(  x\right)
+\zeta^{2}L_{2}\left(  x\right)  \right]  ^{2},
\end{equation}
and%
\begin{align}
V^{\left(  \pm1,m\right)  } &  =%
%TCIMACRO{\dsum \limits_{\mathbf{q}}}%
%BeginExpansion
{\displaystyle\sum\limits_{\mathbf{q}}}
%EndExpansion
\frac{\pi e^{2}}{\epsilon q}L_{m}(2x)e^{-2x}\left[  \frac{1}{2}\zeta^{2}%
x^{3}\right.  \label{Haldane1}\\
&  \left.  +\frac{1}{4}\left(  L_{0}\left(  x\right)  +L_{1}\left(  x\right)
+6\zeta^{2}L_{3}\left(  x\right)  \right)  ^{2}\right]  \nonumber
\end{align}
with $x\mathtt{\equiv}\frac{q^{2}l_{B}^{2}}{2}$ being a dimensionless
variable. From Eq. (\ref{Haldane1}) one can clearly see that the Haldane's
pseudopotentials for LLs $n=1$ and $n=-1$ are still identical in the presence
of the warping term. This is different from the Zeeman splitting effect, which
can induce an asymmetry in the pseudopotentials for $n$=$\pm1$ LLs
\cite{Wang,Zheng}. \begin{figure}[ptbptbptb]
\begin{center}
\includegraphics[width=0.6\linewidth]{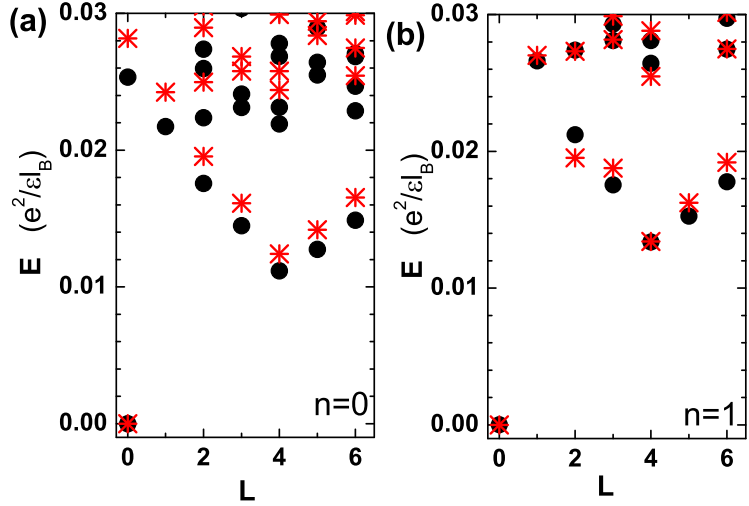}
\end{center}
\caption{(Color online) Exact energies versus the angular momentum $L$ for
$N$=$7$ electrons at $\nu$=$\frac{1}{3}$ TIFQH state. The warping strengh is
chosen as $\zeta$=$0$ (circles) and $\zeta$=$0.2$ (stars). }%
\label{Fig3}%
\end{figure}

Figure \ref{Fig1} plots the Haldane's pseudopotentials of Coulomb interaction
$V^{(n,m)}$ for (a) $n$=$0$ and (b) $n$=$1$ LLs as functions of the relative
angular momentum $m$. Comparing with the results in the absence of the warping
term (black dots in Fig. \ref{Fig1}), one can clearly find that in the
presence of the warping term, the magnitude of the pseudopotentials increases
(red stars in Fig. \ref{Fig1}).

Subsequently, the stability of the TIFQH states should also be modified by the
warping term. The typical results of $V^{(n,1)}/V^{(n,3)}$ ($n$=$0$ and
$n$=$\pm1$) are shown in Fig. \ref{Fig2} as a function of the warping
parameter $\zeta$. One can see that with increasing $\zeta$, $V^{(0,1)}%
/V^{(0,3)}$ increases while $V^{(\pm1,1)}/V^{(\pm1,3)}$ decreases. According
to the composite fermion theory, the larger the value of $V^{(n,1)}/V^{(n,3)}$
is, the more stable the fractional quantum Hall states. Therefore, the
remarkable warping term results in the composite fermions at fractional
filling for LL $n$=$0$ (LLs $n$=$\pm1$) to become more stable (unstable). This
result suggests that on the surface of a TI material with strong (weak)
warping strength, such as Bi$_{2}$Te$_{3}$ (Bi$_{2}$Se$_{3}$), the TIFQH
states for LLs $n$=$0$ ($n$=$\pm1$) may be observed much easier under a
perpendicular magnetic field. We hope this prediction could be detected in
future experiment. \begin{figure}[ptb]
\begin{center}
\includegraphics[width=0.6\linewidth]{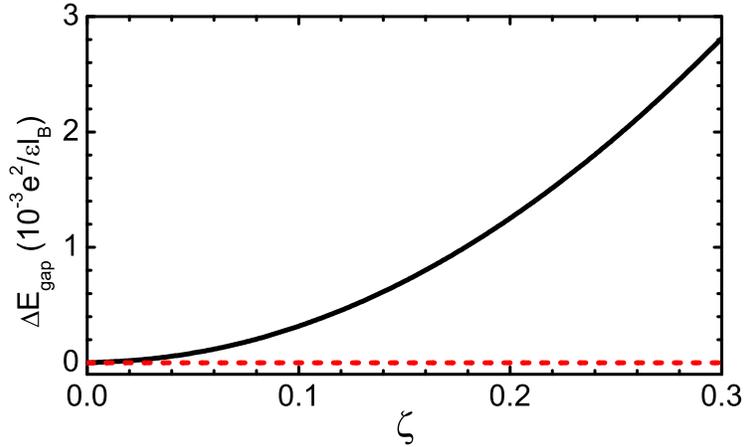}
\end{center}
\caption{(Color online) The increment of the gap width as a \ function of the
warping term $\zeta$ for $N$=$7$ electrons at $\nu$=$\frac{1}{3}$ TIFQH states
for LLs $n$=$0$ (black solid line) and $n$=$\pm1$ (red dashed line). }%
\label{Fig4}%
\end{figure}

In what follows, by using the exact diagonalization method in the spherical
geometry, we investigate the system with the fractional filling factor $\nu
$=$1/(2p+1)$, where $p$ is an integer. For briefness we only illustrate the
$\nu$=$1/3$ TIFQH state, which is realized at $S=\frac{3}{2}\left(
N-1\right)  $ in the spherical geometry with $N$ being the electron number.
Under this configuration, the perpendicular magnetic field is equivalent to a
fictitious radial magnetic field produced by a magnetic monopole at the center
of a sphere of radius $R=\sqrt{S}l_{B}$ (in unit of flux quanta), and the
many-body states could be described by the total angular momentum $L$ and its
$z$ component $L_{z}$.

We show in Fig. \ref{Fig3} the energy spectra of the many-body states at $\nu
$=$1/3$ filling for $N$=$7$ electrons. Comparing the two cases with ($\zeta
$=$0.2$) and without ($\zeta$=$0$) warping term, one can see from Fig.
\ref{Fig3} that the gap width at $n$=$0$ LL between the ground state and the
first excited state has a visible increment while those at $n$=$\pm1$ LL keep
unchanged. Furthermore, we calculate the corresponding excitation gap width
$E_{g}^{n}\left(  \zeta\right)  $ by increasing the warping term from $\zeta
$=$0$ to $\zeta$=$0.3$. The variation $\Delta E_{g}=E_{g}^{n}\left(
\zeta\right)  -$ $E_{g}^{n}\left(  \zeta=0\right)  $ as a function of $\zeta$
are plotted in Fig. \ref{Fig4}, which shows that the TIFQH gap between the
ground state and the lowest excited state at $n$=$0$ LL (solid line) are
sensitively dependent on the warping term while $\Delta E_{g}$ at $n$=$\pm1$
LLs (dashed line) keeps a constant no matter the warping term is included or
not. This also implies that the warping term is different from the Zeeman
splitting and the tilted magnetic field, which will induce a change in the gap
of the TIFQH states at LLs $n$=$\pm1$ \cite{Zheng,Wang}.

In summary, we perturbatively studied the effect of the warping term on the
TIFQH states. It was found that differing from the role of the Zeeman
splitting and the tilted magnetic field, the warping term does not break
symmetry of the Haldane's pseudopotentials for $n$=$\pm1$ LLs. Our results
showed on one hand that, the stability of the $\nu=1/3$ TIFQH states for LL
$n=0$ (LLs $n$=$\pm1$) become stronger (weaker) by increasing the warping
strength of the Fermi surface. On the other hand, the excitation gap for LL
$n$=$0$ increases with increasing the strength of the warping term, while the
gaps for LLs $n$=$\pm1$ are yet insensitive to this term.

This work was supported by Natural Science Foundation of China under Grants
No. 11274049, No. 10904005, No. 11004013, and by the National Basic Research
Program of China (973 Program) under Grant No. 2009CB929103.

\end{document}